\pdfoutput=1

\documentclass{pnastwo}

\def\o{{\scriptsize {\O}}}
\def\footnote{\xdef\@thefnmark{}\@footnotetext}
\usepackage[dvips]{graphicx}
\usepackage{amssymb,amsfonts,amsmath}
\usepackage{amssymb}                 
\usepackage{amsmath}                 
\usepackage{graphicx}                
\usepackage{color}                   

\usepackage{units}                   

\newcommand{\ket}[1]{{\left| #1\right\rangle}}

\newcommand{\var}{\mathrm{var\,}} \newcommand{\cov}{\mathrm{cov\,}}
\newcommand{\intlimits}{\int \limits}

\renewcommand{\Re}{\mathrm{Re\,}} \renewcommand{\Im}{\mathrm{Im\,}}

\usepackage[square, comma, sort&compress]{natbib}
\newcommand{\mycaption}[2][Title]{
  \textsf{\caption{\footnotesize\textbf{#1}.#2}}}

\url{http://arxiv.org/abs/0810.3545}
\issuedate{Issue Date}
\copyrightyear{2008}
\volume{Volume}
\issuenumber{Issue Number}
\footlineauthor{J.Appel et. al}
\begin{document}

\title{Mesoscopic atomic entanglement for precision measurements beyond the standard quantum limit}

\author{J. Appel%
\affil{1}
  {Danish National Research Foundation Center for Quantum
    Optics -- QUANTOP, The Niels Bohr Institute, University of
    Copenhagen, Blegdamsvej~17, DK-2100 K{\o}benhavn~{\O}, Denmark.}, P.~J.~Windpassinger\affil{1}{}, D.~Oblak\affil{1}{}, U.~B. Hoff\affil{1}{}, N.~Kj{\ae}rgaard\affil{1}{}, \and E.~S.~Polzik\affil{1}{} }

\contributor{Submitted to Proceedings of the National Academy of
Sciences of the United States of America}

\maketitle

\begin{article}

\begin{abstract}
  Squeezing of quantum fluctuations by means of entanglement is a well
  recognized goal in the field of quantum information science and
  precision measurements. In particular, squeezing the fluctuations
  via entanglement between two-level atoms can improve the precision
  of sensing, clocks, metrology, and spectroscopy.  Here, we demonstrate
  \unit[3.4]{dB} of metrologically relevant squeezing
  and entanglement for $\gtrsim10^5$ cold caesium atoms
  via a quantum nondemolition (QND)
  measurement on the atom
  clock levels. We show that there is an optimal degree of decoherence
  induced by the quantum measurement which maximizes the generated
  entanglement. A two-color QND scheme used in this paper is shown to have a number of advantages for entanglement generation as compared to a single color QND measurement.
\end{abstract}



\dropcap{W}hen $N_A$ independent two-level atoms are prepared in an
equal coherent superposition of the internal quantum states
$\ket{\uparrow}$ and $\ket{\downarrow}$, a measurement of the
ensemble population difference $\Delta N=N_{\uparrow }
-N_{\downarrow }$ will fluctuate as $\sqrt{N_A}$ (see
Fig.~\ref{fig:one}A). These fluctuations are referred to as
projection noise~\cite{WINELAND1992,Kitagawa1993}.  More generally, the population difference of the
ensemble can be shown to form one component of a collective
pseudo-spin vector~$J$. Taking $J_z=\frac{1}{2}\Delta N$, we have
the variance $(\delta J_z)^2=\frac{1}{4}N_A$ for the case of
independent atoms also referred to as a coherent spin state (CSS).
The CSS minimizes the Heisenberg uncertainty product so that, e.g.,
$(\delta J_z)^2(\delta J_x)^2=\frac{1}{4}|\langle J_y\rangle|^2$ where $\langle J_y\rangle$ is the expectation value of the spin projection operator. At
the expense of an increase in $(\delta J_x)^2$ it is possible to
reduce $(\delta J_z)^2$ (or vice versa) below the projection noise limit while
keeping their product constant.  This constitutes an example of a spin squeezed state (SSS), for which the atoms need to be
correlated.  This correlation is ensured to be non-classical if
\begin{equation}\label{eqsq}
  (\delta J_z)^2<\frac{|\langle J\rangle|^2}{N_A}\Rightarrow
  \xi\equiv\frac{(\delta J_z)^2}{|\langle J\rangle|^2}N_A<1,
\end{equation}
where $\xi$ defines the squeezing parameter. Under this condition
the atoms are entangled ~\cite{Sorensen2001} and the prepared state
improves the signal-to-noise ratio in spectroscopical and
metrological applications ~\cite{WINELAND1992}.

Systems of two to three ions have successfully been used to
demonstrate spectroscopic performance with reduced quantum noise and
entanglement~\cite{Leibfried2004y,Roos2006}. The situation is
somewhat different with macroscopic atomic ensembles where spin
squeezing has been an active area of research in the past
decade~\cite{Kuzmich2000,Hald1999,Orzel2001,Molmer2001,Andre2004,Gerbier2006,Fernholz2008,Takano}.
To our knowledge, no results reporting $\xi<1$ via inter-atomic
entanglement in such ensembles have been reported so far, with a
very recent exception of the paper~\cite{Esteve2008} where
entanglement in an external motional degree of freedom of
$2\cdot10^{3}$ atoms via interactions in a Bose-Einstein condensate
is demonstrated.

\section{Spin squeezing by QND measurements}
\begin{figure*}[t]
  \begin{center}
    \includegraphics[width=0.8\textwidth]{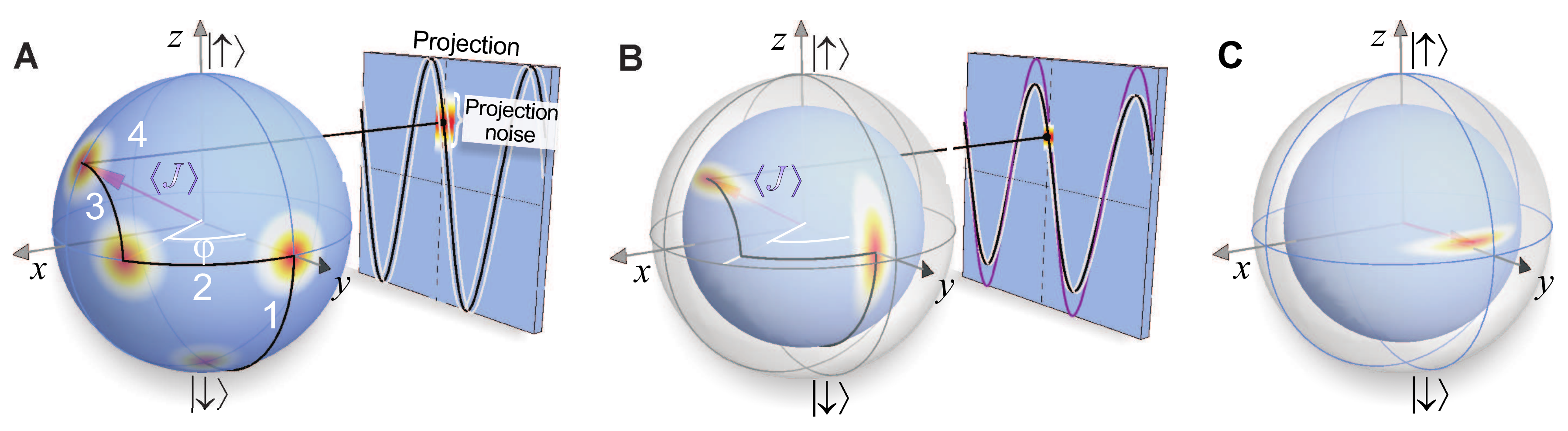}
  \end{center}
  \caption{Ramsey spectroscopy with coherent and squeezed states.
    \textbf{A} - Bloch sphere representation of a Ramsey sequence with
    $N_{A} $ atoms initially in the state $\ket{\downarrow}$ described
    by a Bloch vector $\langle J\rangle$ of length $N_A/2$ pointing
    towards the south pole.  Stage 1: a drive field of frequency $\nu$
    near resonant to the
    $\ket{\downarrow}\leftrightarrow\ket{\uparrow}$ transition
    frequency $\nu_0$ rotates the Bloch vector to the equator
    ($\pi/2$-pulse) bringing each atom in a superposition state
    $\tfrac{1}{\sqrt{2}} \left( \ket{\uparrow} + \ket{\downarrow}
    \right)$.  This CSS is characterized by quantum fluctuations
    $(\delta J_z)^2=(\delta J_x)^2=\frac{1}{4}N_A$ illustrated as a
    fuzzy disk. Stage 2: the atoms evolve freely for a time $T$ and
    the Bloch vector precesses about the z axis by an angle $\varphi
    =2\pi(\nu-\nu_0)T $.  $\varphi $ is recorded after a second $\pi
    /2$-pulse (stage 3) by measuring the population difference (stage
    4): $\cos \varphi =(N_{{ \uparrow} } -N_{{ \downarrow } } )/N_{A}
    $.  Projection noise will result in an uncertainty of the Ramsey
    fringe position. \textbf{B} - The effect of projection noise in
    Ramsey spectroscopy can be reduced by applying an SSS. If the SSS
    is generated via a dispersive QND measurement the accompanying
    spontaneous photon scattering leads to a fraction $\eta$ of
    decohered atoms thereby reducing the radius of the Bloch sphere
    and the Ramsey fringe amplitude by a factor $(1-\eta)$.  Compared
    to the CSS shown in \textbf{A}, the SSS on the reduced Bloch
    sphere \textbf{B} increases the precision in the determination of
    $\varphi$ if the criteria Eq.~\eqref{eqsq} is fulfilled after the
    final $\pi/2$-pulse. \textbf{C} The present paper reports on the
    squeezing of $(\delta J_z)^2$, i.e. along the meridian direction,
    which can be converted into the squeezing in the equatorial
    direction by a microwave pulse. The SSS resides on a Bloch sphere
    of reduced radius as compared to the original CSS.
    \label{fig:one}
  }
\end{figure*}

In this article, we report on the generation of an SSS fulfilling
Eq.~\eqref{eqsq} in an ensemble of $\sim10^5$ atoms via a QND
measurement~\cite{Caves1980,Kuzmich2000,Guerlin2007,Atature2007} of
$J_{z}$. We show how to take advantage of the entanglement in this
mesoscopic system using Ramsey spectroscopy~\cite{WINELAND1992} - one
of the methods of choice for precision measurements of time and
frequency~\cite{Santarelli1999} (Fig.~\ref{fig:one}A). The Figure
presents evolution of the pseudo-spin vector $J$ whose tip is traveling over
 the Bloch sphere. The Ramsey method allows using the
atomic ensemble as a sensor for external fields where the perturbation
of the energy difference between the levels $\Delta
E_{\uparrow\downarrow}$ is measured, or as a clock where the frequency
of an oscillator is locked to the transition frequency between the two
states $\Omega=\Delta E_{\uparrow\downarrow}/\hbar$.
Fig.~\ref{fig:one}B illustrates how a suitable SSS can improve the
precision of the Ramsey measurement provided that the condition of
Eq.~\eqref{eqsq} is fulfilled.

Here we experimentally demonstrate two crucial steps of the
entanglement-assisted Ramsey spectroscopy. First we perform the
projection noise squeezing (Fig.~\ref{fig:one}C) by a QND
measurement. Second, using the Ramsey method, we measure the loss of atomic coherence
$|\langle J\rangle|$ due to QND probing.
Together these results allow us to demonstrate the condition of
Eq.~\eqref{eqsq}. The complete Ramsey sequence can be achieved if
these two steps are supplemented with the rotation of the squeezed
ellipse from the equatorial to the meridian plane (see Fig. S1). We pay particular attention to a reliable
determination of the projection noise level which is an
experimental challenge in its own right.  A QND measurement based on
far-off resonant dispersive probing is always accompanied by
inherent decoherence due to spontaneous emission (shortening of
$|\langle J\rangle|$, as in Fig.~\ref{fig:one}B,C) which affects
entanglement according to Eq.~\eqref{eqsq}. As we show in the paper,
the ensemble
optical depth sets the optimal value of the decoherence which has to be chosen in order to maximize the
generated entanglement.  We employ a dichromatic QND
measurement~\cite{Kuzmich1998} and show that it has several
advantages for the SSS generation~\cite{Saffman2008}.

\subsection{Experimental Setup}
 Cold Cs
atoms are loaded into a far off-resonant optical dipole trap (FORT), aligned to overlap
with the probe arm of a Mach-Zehnder-Interferometer (MZI, shown in Fig.~\ref{fig:two}A).
The FORT is formed by a Gaussian laser beam with a wavelength of
\unit[1032]{nm} and a power of $\unit[2.5]{W}$ focussed to a waist of
$\unit[50]{\mu m}$ inside an evacuated glass cell located in one of
the arms of the MZI (\cite{Oblak2008,Windpassinger2008b}).  Atoms are
loaded into the FORT from a standard magneto-optical trap (MOT)
superimposed onto the FORT, which collects and cools
atoms from the background vapor to $\approx \unit[50]{\mu K}$.  After
loading the FORT, the MOT light is extinguished and a B-field of $\sim
\unit[2]{Gauss}$ is applied, defining a quantization axis orthogonal
to the trapping beam. At this stage the atoms occupy the $(F=4)$
ground level but are distributed amongst the magnetic sublevels. To
polarize the atoms in one of the clock states, a combination of $\pi
$-polarized laser light resonant to the $6S_{1/2} (F=4)\to 6P_{3/2}
(F'=4)$ and $6S_{1/2} (F=3)\to 6P_{3/2} (F'=4)$ transitions is
applied, optically pumping the atoms towards the $(F=4,m_{F} =0)$
state with 80\% efficiency.  Purification of clock state atoms
proceeds by transferring the $(F=4,m_{F} =0)$ state atoms to the
$(F=3,m_{F} =0)$ state using a resonant $\pi$-pulse on the clock
transition and blowing away remaining atoms residing in the $(F=4)$
level with light on the $6S_{1/2} (F=4)\to 6P_{3/2} (F'=5)$ cycling
transition.  The Coherent Spin State preparation is completed by
putting the ensemble in an equal superposition of the clock states
$\bigotimes_{i=1}^{N_A}\left[\frac{1}{\sqrt{2}}
  \bigl(\ket{\downarrow}+\ket{\uparrow}\bigr)\right]_i$ by applying a
resonant $\pi /2$ microwave pulse at the clock frequency. Next we
perform successive QND measurements on the sample, after which all
atoms are pumped into the $F=4$ level to determine the total atom
number $N_A$. The sequence is repeated several thousand times for
various $N_A$. A schematic representation of the experimental sequence
is shown in Fig.~\ref{fig:three}A.

\subsection{Measurement of the projection noise}

\begin{figure}[tbp]
  \begin{center}
    \includegraphics[width=0.7\columnwidth]{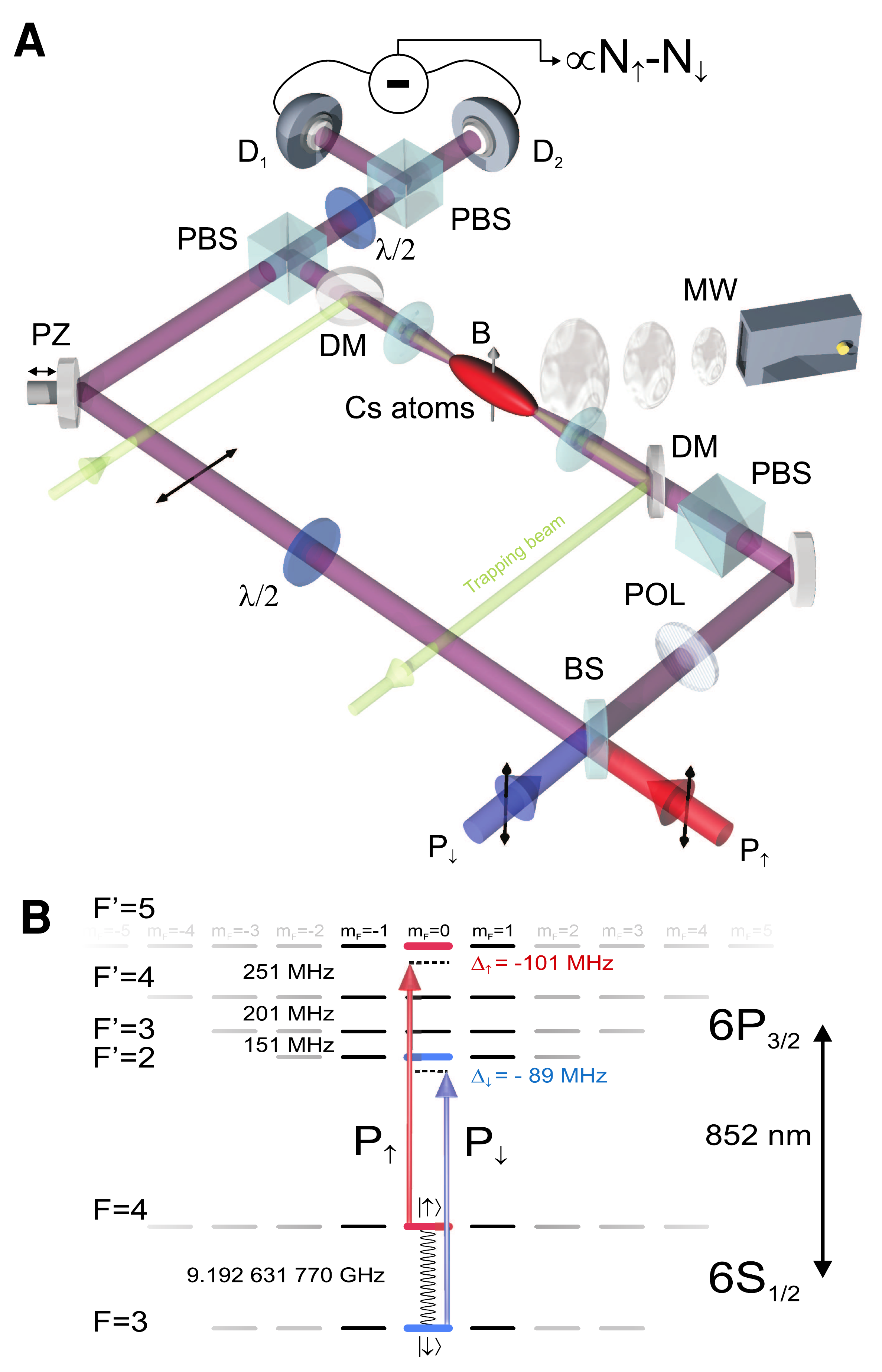}
  \end{center}
  \mycaption[Experimental setup]{\textbf{A} - $\sim10^{5} $ Cs atoms
    cooled to $\sim \unit[50]{\mu K} $ are confined in one arm of a
    MZI by a trapping beam with the waist of $\unit[50]{\mu m}$ folded
    by two dichroic mirrors (DM). The atoms are prepared in a coherent
    superposition of the clock states $\ket{\uparrow}$ and
    $\ket{\downarrow}$, by applying a microwave (MW) $\pi/2$-pulse.
    Two linearly polarized probe beams $P_{\uparrow}$ and
    $P_{\downarrow}$ enter the interferometer via separate ports of
    the input beam splitter (BS). The probes focused to the waist of
    $\unit[27]{\mu m} $ acquire phase shifts proportional to the number of
    atoms in the clock states $N_{\uparrow}$ and $N_{\downarrow}$,
    respectively. An arrangement of polarizers (POL), polarizing beam
    splitters (PBS), and half wave plates ($\lambda/2$) is used to
    adjust the powers and polarizations of the probe and reference
    beams. The combined phase shift
    ($\propto(N_{\uparrow}-N_{\downarrow})$) of the two probes is
    measured in a balanced homodyne configuration.  \textbf{B} -
    Simplified level scheme of Cs showing the D2 line and the
    detunings of the clock state sensitive probes, $P_{\uparrow}$ and
    $P_{\downarrow}$.
    \label{fig:two}
  }
\end{figure}

The dispersive measurement of the clock state population
difference~\cite{Oblak2005,Windpassinger2008a} is realized  by detecting the state dependent phase shift of off-resonant probe laser light.  In the present experiment
one probe $P_\downarrow$ is coupled to the state $\ket{\downarrow}
\equiv 6 S_{1/2}(F=3,m_F=0 )$, while a second probe $P_\uparrow$ is
coupled to the state $\ket{\uparrow} \equiv 6 S_{1/2}(F=4, m_F=0) $
(see Fig.~\ref{fig:two}B).

The two probe beams enter the interferometer through different
ports, so that the phase shifts imprinted on them by the atoms
contribute with opposite signs to the differential signal $\Delta n$
from the detectors $D_1,D_2$. As discussed below this geometry
together with a suitable choice of probe detunings provides
compensation of a deleterious probe-induced shift of the measured
frequency $\Omega$. Denoting the sum photosignal as $n$ we define
\begin{equation}
  \label{eq:ja_1}
  \phi = \frac{\Delta n}{n} = \frac{\delta n}{n} + k_\uparrow N_\uparrow - k_\downarrow
  N_\downarrow,
\end{equation}
where $\delta n=\delta n_\uparrow+\delta n_\downarrow$ denotes the
total shot noise contribution of both probe colors. The probes are
of equal intensity and their detunings are chosen such that the
coupling constants $k \equiv k_\uparrow=k_\downarrow$ are equal. Hence the
normalized differential signal $\phi$ yields a $J_z$ measurement
with the optical shot noise added:
\begin{eqnarray}
  \label{eq:ja_2}
  \phi   & = &  \frac{\delta n}{n} + k\Delta N = \frac{\delta n}{n} + 2k J_z,\\
  \var(\phi) & =&  \frac{1}{n} + k^2 \var(\Delta N). \label{eq:ja_3}
\end{eqnarray}
For atoms in a CSS we have $\var(\Delta N)=N_A$ and
Eq.~\eqref{eq:ja_3} predicts a \emph{linear} increase of the
projection noise with the number of atoms.

Figure~\ref{fig:four} shows that we have observed the projection noise of atoms
as evidenced by the almost perfect linear fit to the noise data
(blue points). To further confirm
that the linear part of Fig.~\ref{fig:four} is the quantum
projection noise we verify the value of $k^2$ for the inferred
projection noise slope by independent means (see Supplementary
information). Achieving this linearity is a demanding experimental
task because it requires technical noise, e.g., fluctuations
of the probe power and the interferometer length to affect $\Delta
N$ measurements well below the level of $1/\sqrt{N_A} \approx 2\cdot
10^{-3}$ over the time scale of the experiment.
\subsection{Conditional noise reduction by QND measurement}
The ability to measure the atomic spin projection with a sensitivity
limited by the shot noise of light allows us to produce a
conditionally spin squeezed atomic state.
After preparation of a CSS, we use $n_1$ photons to measure $J_z$ as
outlined above and obtain a measurement result $\phi_1$, which is
randomly distributed around zero with a variance $\frac{1}{n_1} +
k^2 N_A$ (see Eq.~\eqref{eq:ja_3}, blue dots in
Fig.~\ref{fig:four}).

\begin{figure}[tbp]
  \begin{center}
    \includegraphics[width=0.7\columnwidth]{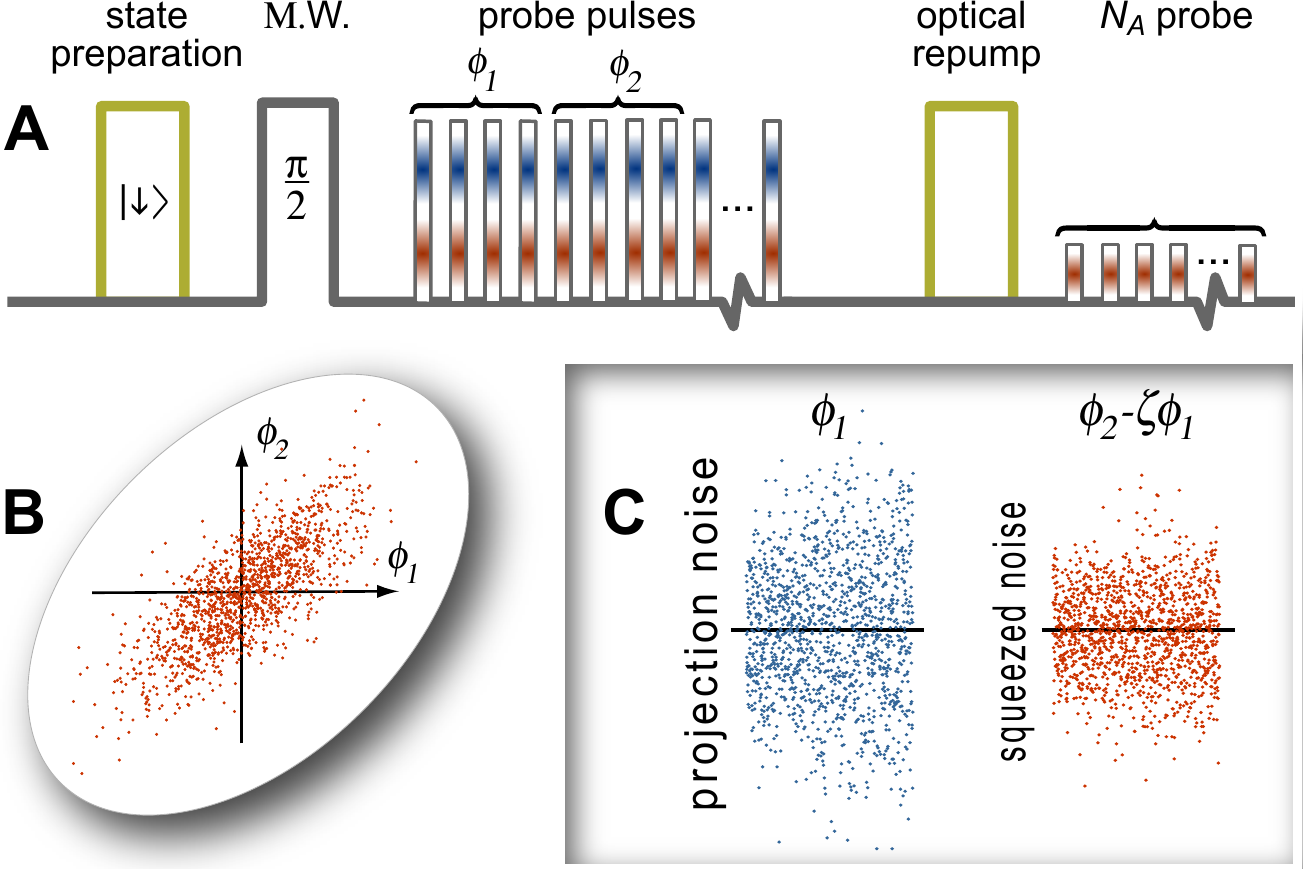}
  \end{center}
  \mycaption[Pulse sequence and noise data]{ \textbf{A} - Atoms are
prepared in state $\ket{\downarrow}$ by an optical pumping sequence
and then rotated to the superposition state $\frac{1}{\sqrt{2} } \left
( \ket{\uparrow} + \ket{\downarrow} \right)$ by a microwave $\pi /2$
pulse before the train of 10 probe pulses is applied. Combining the results of several pulses we can change the effective QND measurement strength as explained in the text. The first
effective probe pulse measurement result $\phi_{1} $ yields the
statistics of the $J_{z} $ for the CSS. The second effective pulse
measurement result $\phi_{2} $ verifies the squeezing provided it
is sufficiently correlated with $\phi_{1} $. $N_A$ is measured at the
end of each sequence. \textbf{B} - Correlations between the first and
the second pulse measurements. \textbf{C} - the projection noise
manifested in the random scattering of about 2000 measurements of $
\phi_{1} $; and the spin squeezed state displayed as the reduced
noise in $\phi_{2} $ when the QND result is used as $\left(\phi_{2} -
\zeta\phi_{1} \right)$
\label{fig:three}
  }
\end{figure}

Using the information obtained in the first measurement we can, to a
certain degree, predict the outcome $\phi_2$ of a successive $J_z$
measurement performed on the same ensemble of atoms. The best
estimate for $\phi_2$ is $\zeta \phi_1$, which results in a
conditionally reduced variance
\begin{equation}
  \label{eq:1}
  \var\left(\phi_2 - \zeta \phi_1\right) =\frac{1}{n_2} + \frac{1}{1+\kappa^2} k^2
  N_A,
\end{equation}
that displays a reduction of the projection noise by
$\frac{1}{1+\kappa^2}$ (cf. red diamonds in Fig.~\ref{fig:four}).
The measurement strength $\kappa^2= n_1 k^2 N_A$ describes the ratio
of the atomic noise to the shot noise of light and $\zeta =
\cov(\phi_1,\phi_2) /\var(\phi_1) = \kappa^2/(1+\kappa^2)$. A
QND~measurement with a finite strength $\kappa^2$ leads to a finite
correlation between the two measurements, as shown in
Fig.~\ref{fig:three}B. For $\kappa^2=3.2$ observed for $N_A=1.2
\cdot 10^5$ atoms we expect a conditionally reduced variance of
$\unit[-6.2]{dB}$ with respect to the projection noise level. In the
experiment we find the value $\unit[-(5.3 \pm 0.7)]{dB}$ as shown in
Fig.~\ref{fig:four}.
\subsection{Decoherence}
Spontaneous emission cau\-sed by the QND probes is a fundamental irreversible decoherence mechanism which
affects the SSS in two ways. First, it can change the value of
$J_{z}$ by redistributing atomic populations via inelastic Raman scattering. This effect would be
particularly important if a single QND probe coupled to both clock
levels were used. Except for special cases such as discussed in ~\cite{Ozeri}, the single color probing causes Raman scattering between the clock levels. With the two-color QND scheme which we use here
the population redistribution between the hyperfine levels is
practically absent because of the selection rules and the choice of detunings (atoms in the ground state level $F=4$ are predominantly excited to the level $F'=5$ and hence cannot decay to the ground state $F=3$, and similarly for the reverse scattering). The effect of the redistribution between the
magnetic sublevels within a given hyperfine level on the projection
noise squeezing is very small (estimated to be less than $1\%$), as
also proven by a good agreement between the observed and predicted
degree of this squeezing (see the section on dichromatic probing and
supplementary information for further discussion of this issue).

The second and main effect of the spontaneous decoherence on
entanglement is due to the reduction of the coherence between the
clock levels. Both inelastic Raman and elastic Rayleigh spontaneously scattered photons lead to shortening of
the mean collective spin vector $|\langle J\rangle| \rightarrow
(1-\eta)|\langle J\rangle|$ and hence to the reduction in Ramsey
fringe amplitude (see Fig.~\ref{fig:one}B). The degree of spin
squeezing as defined by Eq.~\eqref{eqsq} depends on the fraction
$\eta$ of atoms which decohere as a result of spontaneous photon
scattering during dispersive QND probing.  The QND measurement
strength can be cast as $\kappa^2\propto d\eta$ where $d$ is the
resonant optical depth of the sample~\cite{Hammerer2008}.  This
highlights the trade-off between information gained through strong
coupling and coherence lost due to spontaneous emission.

An inhomogeneous AC Stark shift due to the transverse intensity profile of the probe beam can, in principle, cause an additional dephasing and decoherence. However, as discussed in the section on dichromatic QND, this deleterious effect is strongly suppressed in our QND sequence.

We determine $\eta$ via the reduction in the Ramsey fringe amplitude in a
separate echo spectroscopy
experiment~\cite{Oblak2008,Windpassinger2008b}. The technique and
results are summarized in Fig.~\ref{fig:etaestim}. The reduction in echo fringe amplitude as
a result of the probe light thus provides an upper bound for the
decoherence inflicted.

\subsection{Squeezing and entanglement}
The noise measurement data presented in Fig.~\ref{fig:four}
corresponds to $\eta=\unit[20]{\%}$ as measured in echo spectroscopy.
According to Eq.~\ref{eqsq} metrologically relevant spin squeezing and entanglement $\xi<1$
for a given $N_A$ can be claimed if the
conditionally reduced variance of the verification measurement (red
diamonds in Fig.~\ref{fig:four}) is less than the projection noise
scaled down by the factor $(1-\eta )^{2}$ (dash-dotted line in
Fig.~\ref{fig:four}). In the inset of Fig.~\ref{fig:four} we
consider the maximum $N_A$-bin of the data and plot $\xi$ versus
$\eta$, varying the probe photon number by combining consecutive
probe pulses (see Methods for details of the data analysis). Maximum squeezing $\xi =\unit[-(3.4\pm
0.7)]{dB}$ is observed with $\eta=\unit[20]{\%}$, corresponding to
probing the atoms with $1.3\cdot 10^{7} $ photons.  The squeezing
reduces as $\eta$ increases further, confirming the notion that
though a stronger measurement enables more precise estimation of
$J_z$, this reduction in spin noise eventually ceases to be
spectroscopically relevant as a result of decoherence.

\subsection{Dichromatic QND}
A single probe introduces a phase shift of the Ramsey fringe of approximately $0.8$ radians due to the Stark shift. If two properly balanced probes $P_\downarrow$ and $P_\uparrow$ have the same spatial mode the differential
AC-Stark shift of the clock states $\ket{\downarrow}$ and
$\ket{\uparrow}$ is strongly suppressed and the phase shift of the Ramsey fringe is reduced below the measurement precision. However, even in the presence of inevitable probe fluctuations which lead to a differential Stark shift, this shift does not affect the QND Ramsey sequence. The differential Stark shift caused by the misbalance of the two probes would affect the initial value of $J_{x}$, that is it would add noise to the anti-squeezed component of the spin without affecting the squeezed component. $J_{x}$ is not measured in the Ramsey sequence (see eq. S8 in the Supplementary information), and
therefore the dichromatic QND method
does not lead to a differential light shift of the clock
transition~\cite{Saffman2008}. This feature is most important for both the
precision and the accuracy in metrological applications.

Another advantage of the two-color probing is due to the fact that, as
opposed to the single probe QND schemes~\cite{Oblak2005} where
detuning is fixed roughly in the middle between the atomic levels of
interest, the detuning of two probes can be adjusted. The detuning is
then chosen so that the optimal value of the parameter $\eta$ is
obtained for a photon number around $10^{7}$, a convenient value which
minimizes the combined effect of the laser intensity and frequency
classical fluctuations.  At the same time for a single probe QND the
photon number corresponding to the optimal $\eta$ would be
approximately $10^{10}$, thus demanding a challenging level of
stabilization of the probe power to much better than $10^{-5}$.

Finally, we discuss the fundamental advantage of a dichomatic QND
measurement with cyclic transitions~\cite{Saffman2008}. Such probing does not add any
noise to $\delta J_{z}^{2}$. Hence, from
Eq.~\eqref{eqsq} together with the theoretical estimate for $\delta
J_{z}^{2}$ after the QND measurement, an optimal squeezing of
$\xi_{\min } =\frac{9}{4} \frac{1}{1+\kappa ^{2} } \propto
\frac{1}{d} $ is predicted for $\eta =\frac{1}{3}$, assuming a large
resonant optical depth $d$. Note that the optimal experimental value
of $\eta$ as shown in the inset of Fig.~\ref{fig:four} is close to,
albeit somewhat smaller than, the theoretical optimum.

We emphasize that the theoretical scaling of $\xi _{\min } \propto
d^{-1} $ for the dichromatic QND method using cyclic transitions is
more favorable than the scaling $\xi _{\min } \propto d^{-1/2} $ for
a conventional single-color QND scheme with
cross-pumping~\cite{Hammerer2008}.

The probe transitions used in the present paper are not exactly
cyclic. The spontaneous emission does not reshuffle atoms between
the hyperfine levels but it does cause Raman scattering
$m_{F}=0\rightarrow m_{F}=\pm1$ within each hyperfine level.
Atoms spontaneously scattered to both $m_{F}=0$ and $m_{F}\neq0$ do not add noise
to the QND-assisted Ramsey sequence (see Supplementary information
for details). The only additional partition noise of $J_{z}$ (less
than $1\%$) which appears after the spontaneous emission is due to
slightly different probe coupling to atoms $m_{F}=0$ and
$m_{F}\neq0$.
Interestingly, it becomes obvious from this discussion that the
atoms which decohered to $m_{F}\neq0$ states are part of the
entangled state since if they were removed after the QND measurement
the parameter $\xi$ would grow due to the increased uncertainty of
$\delta J_{z}^{2}$.

\subsection{Conclusion and Outlook}
In summary we have demonstrated a reduction of projection noise to
$\unit[-(5.3\pm 0.6)]{dB}$ and metrologically relevant spin squeezing
and entanglement of $\unit[-(3.4\pm 0.7)]{dB}$ on the Cs microwave
clock transition.  We analyze in detail the role of spontaneous
photon scattering in the Ramsey sequence combined with QND measurements and
demonstrate the optimal balance between decoherence and the
measurement strength for generation of spin squeezing. The
measurement precision improves with the number of atoms, hence it is
important that we have demonstrated entanglement with the largest
to-date number (over $10^{5} $) of cold and trapped atoms. We show
that the dichromatic QND measurement has several advantageous
features in the clock and Ramsey spectroscopy settings.

The phase noise of our microwave reference (\textit{HP8341B})
prevented us from a demonstration of the effect of the spin
squeezing in a full clock cycle, even for integration times $<
\unit[10^{-4}]{s}$. We expect to be able to improve on that with a
better microwave source.

A crucial parameter defining the degree of spin squeezing generated
with the QND method is the resonant optical depth of the atomic
ensemble. Here we have shown that significant entanglement can be
obtained with a modest optical depth of $\sim16$. The requirement of
high optical depth poses a limitation on the applicability of our
results to, e.g., projection noise limited fountain
clocks~\cite{Santarelli1999} where the optical depth is
usually kept around unity in order to avoid collisional shifts.

Spin squeezing is on the current agenda for optical clocks
\cite{Meiser2008,Wilpers:2002,Ludlow:2008y} which are affected by
collisions to a much smaller extent. The first step along the lines discussed in the present paper in state-of-the-art lattice clocks has been demonstrated very recently in ~\cite{Lodewyck} where a nondestructive measurement using a Mach-Zehnder interferometer with
sensitivity near the projection noise level has been reported.  The dichromatic QND method can be
particularly convenient in optical clocks if both clock levels are
probed, because the levels separated by optical energy can be probed
with two independent lasers.

In systems with a
low single pass optical depth the effective optical depth can
be increased by utilizing an optical cavity as demonstrated recently
~\cite{SchleierSmith2008} where results similar to ours have been
obtained. Sensing and metrology for which a high
optical depth and associated collisions are not a problem
will probably be the main field of application for this method,
unless the collisional shift itself is the goal of a Ramsey
experiment~\cite{Hart:2007}.
\newenvironment{methods}[1][Methods]{ \vskip6pt%
  \parskip=8pt%
  \baselineskip=9pt%
  \materialfont%
  \def\textit##1{{\materialitfont ##1}}%
  \def\it{\materialitfont}%
  \def\bf{\materialbffont}%
  \def\textbf##1{{\materialbffont ##1}}%
  \def\section##1{\noindent{\materialbffont ##1. }}%
  \def\subsection##1{\noindent{\materialbffont ##1. }}%
  \noindent {\subsectionfont #1}\\[2pt]}{\vskip1sp}%

\begin{methods}

  \subsection{Noise Measurement Sequence}
  The clock state sensitive probes $P_{\uparrow}$ and $P_{\downarrow}$
  enter the interferometer via opposite input
  ports (see Fig.~\ref{fig:two}) and are phase locked to achieve
  common mode rejection of frequency fluctuations when probing the
  atoms.  Because of the opposite inputs, the differential output
  signal $D_2-D_1$ for the empty interferometer has opposite signs for
  $P_{\uparrow } $ and $P_{\downarrow } $, and their intensity ratio
  is
  controlled as to achieve zero mean of the signal; hence the
  dichromatic interferometer becomes nearly insensitive to geometric length
  variations. With the intensity ratio fixed the detunings of
  $P_{\uparrow } $ and $P_{\downarrow } $ are fine-tuned to achieve
  zero mean signal for atoms in the coherent
  superposition. The two probes are mode matched,
  so that the transverse spatial distributions of $P_\uparrow$ and
  $P_\downarrow$ (waist size $\unit[27]{\mu m}$) are very similar.
  Hence the differential AC Stark shift on the clock transition from a
  dichromatic probe pulse is significantly reduced across the entire
  atomic volume.

  The geometric path length difference between the probe and reference
  arms of the interferometer is controlled by a piezo actuated mirror.
  Prior to atomic probing the mirror position is adjusted in a
  feedback loop to achieve zero differential interferometer output
  near the white light position using a sequence of weak auxiliary laser
  pulses at \unit[840]{nm}.  This light virtually does not interact
  with the atoms and provides a reference position from which the
  mirror is subsequently offset to obtain an \emph{individually}
 balanced output for each of the two probe beams. This allows to reject
  intensity noise in the probe pulses during superposition state measurements.  The offset value depends on
  the (expected) number of trapped atoms.  The atomic ensemble is
  probed at $\unit[20]{\mu s}$ intervals using a sequence of 20
  dichromatic pulses with a duration of $\unit[10]{\mu s}$ and
  $n_\text{pulse}$ photons per color, per pulse ($1.83 \cdot 10^6$ in
  the probe arm). The differential interferometer signal is sampled on
  a digital storage oscilloscope from which we determine the
  differential photon numbers $\{ p_{1} ,p_{2} ...p_{20} \}$.  After
  probing the superposition state the interferometer is reset to its
  reference position and the atoms are optically pumped into the
  $(F=4)$-level to determine $N_A$. The trapped atoms are recycled for
  additional measurements by re-initializing the ensemble using optical pumping
  and purification as described above. During each MOT cycle four
  ensembles are interrogated and additionally three reference
  measurements without atoms are recorded.

  \subsection{Data Analysis}
  We combine up to $P=10$ successive pulses which allows us to vary the effective
  QND probe photon number $n=2 P n_\text{pulse}$, and obtain the phase shifts
  $\phi_1=(p_1 + \ldots p_P)/n$, $\phi_2=(p_{P+1} + \ldots +
  p_{2P})/n$. These measurements are sorted and grouped according to
  the atom number $N_A$. We fit a general quadratic function
  $V(N_A)=v_0 + v_1 N_A + v_2 N_A^2$ (solid blue line in
  Fig.~\ref{fig:four}) to $\var(\phi_1)$ and $\var(\phi_2)$ (blue
  points and stars) and another function $C(N_A)=c_0+c_1 N_A +c_2
  N_A^2$ to $\cov(\phi_1,\phi_2)$ to identify the light noise
  contribution $v_0-|c_0|-d_0$ (light blue area), $d_0=$ detector
  noise (dark blue area), and the projection noise $v_1 N_A$ (green
  field, dashed line). Knowing the outcome of a $\phi_1$
  measurement the best prediction for the successive $\phi_2$
  measurement on the same sample is $\zeta \phi_1$, with
  $\zeta=\cov(\phi_1,\phi_2)/\var(\phi_1)$. The conditionally reduced
  variance $\var(\phi_2 - \zeta \phi_1)$ (red diamonds) is predicted
  by $R(N)=V(N)\left(1- \bigl(\tfrac{C(N)}{V(N)}\bigr)^2\right)$ (red
  line).  From a spin-echo measurement with pulses of $n_\text{se}=7.4
  \cdot 10^6$ photons altogether resulting in a fringe contrast
  reduction of $\eta_\text{se}=\unit[11]{\%}$ the value of spin
  squeezing for a QND measurement employing a total of
  $n_\text{Probe}$ photons in the probe arm is then calculated as
  $\text{SQ}=10 \log_{10}\left(\frac{R(N_{A,\text{max}})-R(0)}{v_1
      N_{A,\text{max}}} (1-\eta_\text{se})^{-2 n_\text{Probe} /
      n_\text{se}}\right)$.  To compensate for slow drifts in our
  setup, from each set of raw data $p_{1}\ldots p_{20}$ we subtract
  the individual pulse data recorded in the previous MOT cycle and
  perform our data analysis on the differential values.

  \subsection{Decoherence measurement}
  \begin{figure}[tbp]
  \begin{center}
    \includegraphics[keepaspectratio,width=0.9\columnwidth]{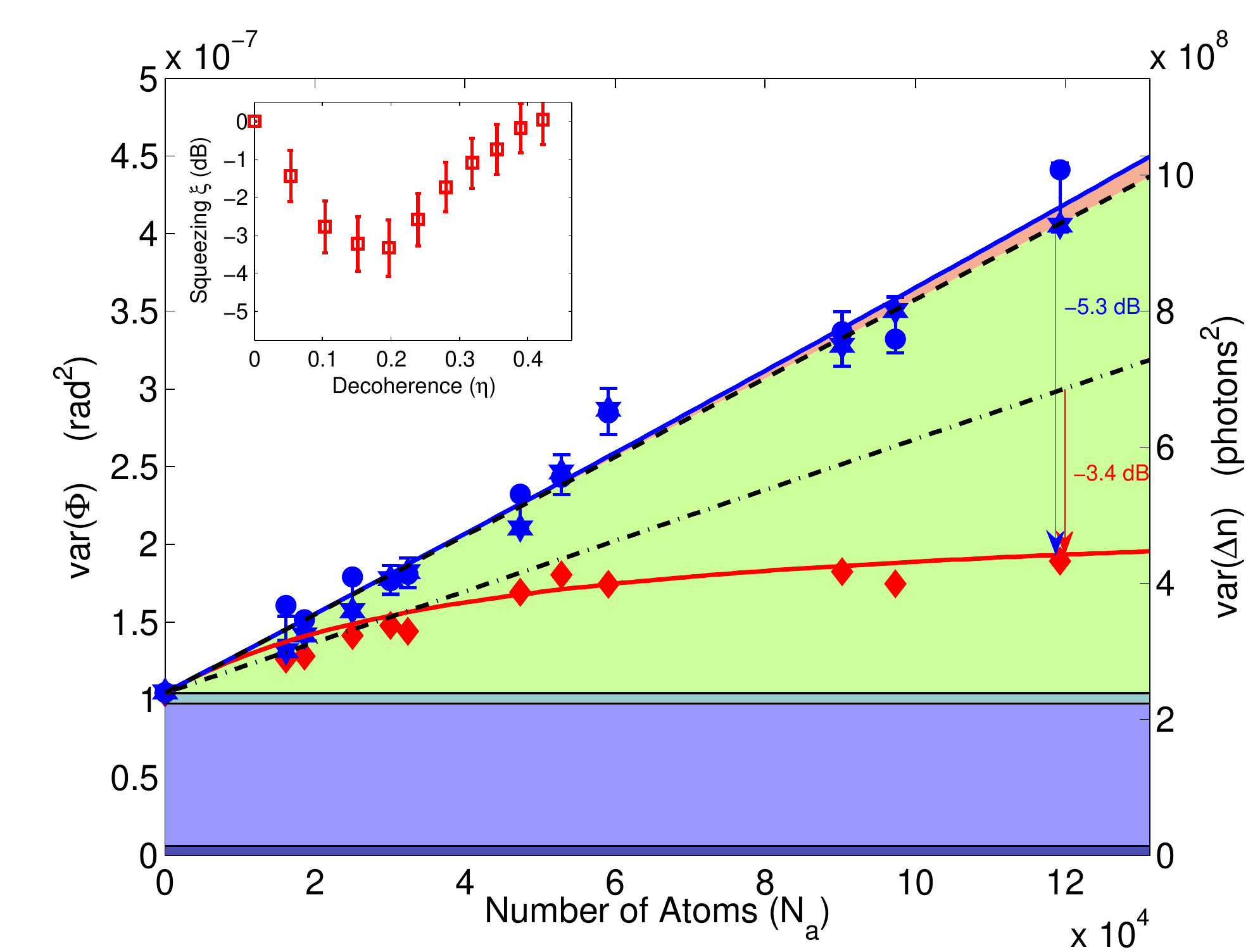}
  \end{center}
  \mycaption[Projection noise and spin squeezing]{ Blue points, stars:
    Variances $\var(\phi_1)$, $\var(\phi_2)$ of the $J_z$ spin noise
    measurement of atoms in a CSS versus $N_A$, error bars:
    corresponding statistical uncertainty, centered on
    $\tfrac{1}{2}(\var(\phi_1)+\var(\phi_2))$; solid blue line:
    quadratic fit (see Methods section).  Dashed line: CSS projection
    noise.  Dash-dotted line: equivalent CSS projection noise reduced
    by the loss of atomic coherence. Red diamonds: Conditionally
    reduced variance of a second $J_z$ spin measurement predicted by
    the first: $\var(\phi_2 - \zeta \phi_1)$.  Red line: reduced noise
    of SSS predicted from quadratic fits to projection noise data(see
    Methods section).  According to the scaling behavior we classify
    different noise contributions (see data analysis): Classical
    fluctuations are represented by the cyan (empty interferometer)
    and red area (atom-light interaction related).  Blue areas:
    optical shot noise (light blue) and detector noise (dark blue).
    Green area: projection noise.  \textbf{Inset}, Spin squeezing
    $\xi$ (red boxes) as a function of the decoherence parameter
    $\eta$, corresponding to the fits (dash-dotted black and red line)
    evaluated for the atom number of the rightmost bin; error bars:
    standard deviation over analysis runs binning into 5 to 30 groups
    with respect to $N_A$.
    \label{fig:four}
  }
\end{figure}
\begin{figure}[tbp]
  \begin{center}
    \includegraphics[width=0.65\columnwidth]{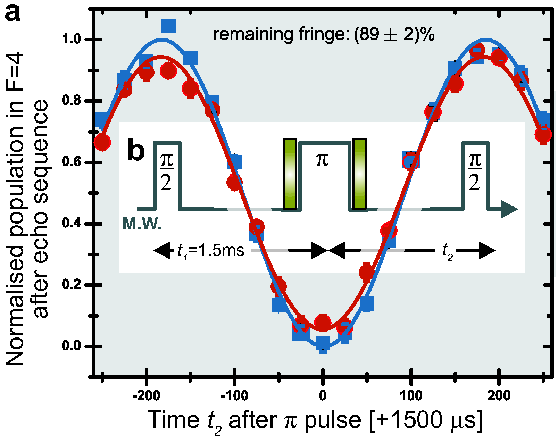}
  \end{center}
  \mycaption[Decoherence measurement]{(A) Sample traces of the
    $F=4$ population as function of the second $\pi/2$-pulse delay
    $t_2$.  Blue squares serve as reference level for the fringe
    amplitude without any probe light pulses. Red points were obtained
    with a combined photon number in the two pulses of $2n_{p}
    =7.4\cdot 10^{6}$. From the fitted curves the reduction of the
    fringe amplitude is inferred giving an upper bound of $\eta
    =(0.11\pm 0.02)$. (B) Experimental sequence similar to
    that of Fig~1a) with addition of a microwave $\pi$-pulse to
    reverse dephasing. Without the probe light pulses, the microwave
    $(\pi/2,\pi,\pi/2)$--pulse sequence compensates all reversible
    dephasing effects.  Absorption of light from the two probe pulses
    causes a fraction $\eta$ of atoms to undergo (irreversible)
    spontaneous emission.  \label{fig:etaestim} }
\end{figure}

  In Fig.~\ref{fig:etaestim} we show the decoherence measurements for
  two dichromatic probe light pulses containing in total $7.4 \cdot
  10^{6}$ photons.  The data has been taken at a microwave detuning of
  $\Delta\Omega=\unit[3]{kHz}$ from the clock transition frequency and
  the inversion of the phase through the microwave $\pi$-pulse was
  induced after \unit[1.5]{ms} of free evolution.  The two light
  pulses reduce the echo-fringe by $\unit[11\pm 2]{\%}$ and thus this
  forms an upper bound for $\eta$ for the combined effect of these
  probe pulses.  From this we can deduce the decoherence $\eta$
  corresponding to a probe pulse with a photon number $n_P$ as
  $\eta(n_P) \approx 1 - (1-0.11) ^ \frac{n_P}{7.4 \cdot 10^6}.$

  We have also verified that the dichromatic probing employed in the
  QND measurement indeed provides a significant reduction of the
  differential Stark shift by measuring the fringe visibility
  reduction in the absence of the $\pi$-pulse. In cases when the two
  probes were mode matched to better than \unit[97]{\%} we indeed
  observed an identical fringe reduction irrespectively of the
  $\pi$-pulse.
\end{methods}

\appendix[Derivation of the QND measurement equation]
To describe the propagation of the probe light through the atomic
gas, we express its complex valued susceptibility by
\begin{equation}
  \label{eq:2}
  \chi =\frac{n_A}{L} \frac{\lambda}{ 2 \pi} Q, \qquad Q=
  - \frac{3 \lambda^2}{4 \pi} \sum_l \frac{\wp_l}{\Delta_l/\gamma + i/2 },
\end{equation}
where for each probe color the sum is taken over all possible
transitions; $\wp_l$ denote Clebsch-Gordan
coefficients, $\Delta_l$ is the detuning from resonance, $n_A$ is the
atom column density, $L$ is the interaction length, $\lambda$ is the vacuum
wavelength, and $1/\gamma$ the excited state lifetime.

Since the $P_\uparrow$ ($P_\downarrow$) probe only interacts
with atoms in the $\ket{\uparrow}$ ($\ket{\downarrow}$) state we can
calculate the phase shift $\theta$ and absorption $e^{-\alpha}$ for a plane wave of the probe light passing through
the cloud as follows:
\begin{eqnarray}
  \label{eq:3}
  \theta_{\uparrow,\downarrow}  =
  & \frac{1}{2} \Re \frac{2 \pi}{\lambda} L \chi_{\uparrow,\downarrow}
  &  =  \frac{1}{2} n_{A\uparrow,\downarrow} \Re Q_{\uparrow,\downarrow}, \\
  \alpha_{\uparrow,\downarrow} =
  & \Im \frac{2 \pi}{\lambda} L \chi_{\uparrow,\downarrow}
  & =  n_{A\uparrow,\downarrow} \Im Q_{\uparrow,\downarrow}.
\end{eqnarray}

The Gaussian spatial profiles of the two probes $I_{P}
(r)=\frac{2}{\pi w^{2} } e^{-2\frac{r^{2}}{w^{2}}}$ with a waist size
$w=\unit[27]{\mu m}$ are mode matched to better than $97\%$ by
maximizing the interference fringe between the two spatial modes using
the same laser as input.  The atomic cloud is laser cooled and trapped
in a far detuned dipole trap, as described in detail
in~\cite{Oblak2008}, so that the transverse size of the cloud is about
a factor of two larger than the probe cross section, hence the total
column density $n_{A}$ is considered constant across the probe cross
section.

$n_{A}$ can be expressed as the \emph{sum} of the column densities
of the individual states:
\begin{equation}
  \label{eq:12}
  n_{A}=n_{A\uparrow}(r,\vartheta)+n_{A\downarrow}(r,\vartheta)
\end{equation}
Due to quantum fluctuations in the \emph{difference} of the
populations in the $\ket{\uparrow}$ and $\ket{\downarrow}$ states,
the individual contributions $n_{A\uparrow,\downarrow}(r,\vartheta)$
can differ for different positions $(r,\vartheta)$ within the beam.

The measurement of the probe phase shift is performed by
interference with a mode matched reference beam and hence the
differential number of photons measured by the detectors $D_1,D_2$
is given by
\begin{gather}
  \label{eq:4}
  \Delta n_{\uparrow ,\downarrow } = \pm 2\sqrt{tn_{R\uparrow
      ,\downarrow } n_{P\uparrow ,\downarrow } } \theta_{\uparrow
    ,\downarrow }, \\
  \label{eq:6}
  \text{where} \quad \theta_{\uparrow ,\downarrow } = \frac{1}{2} \Re
  Q_{\uparrow,\downarrow} \intlimits_0^{2 \pi} \intlimits_0^\infty
  \mathrm{d}\vartheta r\mathrm{d}r \, I(r)
  n_{A\uparrow,\downarrow}(r,\vartheta)
\end{gather} is the spatially averaged phase shift of the probe(s),
$t=\unit[63]{\%}$ is the probability of the homodyne detection of a
photon which has interacted with the clock atoms, $n_{R} $ is the
number of detected photons from the reference arm of the
interferometer, $n_{P} $ is the number of probe photons interacting
with atoms, and $n_{A\uparrow ,\downarrow } $ is the column density
of the atomic cloud in the two clock states respectively.  Since the
probes are injected via opposite input ports, the detector signals
have opposite signs. In deriving Eq.~\eqref{eq:6} we assumed that
the phase shifts $\theta_{\uparrow,\downarrow}$ are small such that
$\sin(\theta_{\uparrow,\downarrow}) \approx
\theta_{\uparrow,\downarrow}$.

The photon numbers for both probes are matched ($n_R \equiv
n_{R\uparrow}=n_{R\downarrow}$, $n_P \equiv
n_{P\uparrow}=n_{P\downarrow}$) whereas the detunings are chosen
such that the differential number of photoelectrons (the signal) for
the two probes is zero for equal populations of the clock levels,
i.e., $\Re Q_\uparrow = \Re Q_\downarrow$. Hence with $\tilde n
\equiv \sqrt{t n_{R} n_{P}} Q$ denoting the interferometer fringe
amplitude the total signal due to atoms can be expressed as
\begin{equation}
  \label{eq:5}
  \Delta n= \tilde n \left[\Re Q
    \intlimits_0^{2 \pi} \intlimits_0^\infty \mathrm{d}\vartheta r
    \mathrm{d}r \, I_P(r) \Bigl(
    n_{A\uparrow}(r,\vartheta) -n_{A\downarrow}(r,\vartheta)\Bigr)
  \right].
\end{equation}

When the atoms are prepared in one of the clock states, say state
$\ket{\uparrow}$, the interferometer signal measures the total
number of probed atoms $ \Delta n_{\ket{\uparrow}} = n_{A\uparrow}
\tilde n \Re Q$.

When the clock atoms are prepared in a coherent superposition of the
states $\ket{\uparrow}$ and $\ket{\downarrow}$ (the CSS), the
expectation value for the signal is zero: $\langle \Delta
n_\text{CSS}\rangle =0$.  Since the CSS is a product state there is
no correlation in the noise of $n_{A\uparrow}-n_{A\downarrow}$ at
different positions within the beam and therefore (omitting optical
shot noise for the rest of this section):
\begin{align}
  \label{eq:9}
  & \var(\Delta n_\text{CSS})  \nonumber \\
  & = \tilde n^2 (\Re Q)^2 \var \left[ \intlimits_0^{2 \pi}
    \intlimits_0^\infty \mathrm{d}\vartheta r \mathrm{d}r \, I_{P}(r)
    \bigl( n_{A\uparrow}(r,\vartheta)
    -n_{A\downarrow}(r,\vartheta)\bigr)
  \right] \nonumber \\
  & = \tilde n^2 (\Re Q)^2 \intlimits_0^{2 \pi} \intlimits_0^\infty
  \mathrm{d}\vartheta r \mathrm{d}r \, \bigl(I_{P}(r)\bigr)^2 \!\!
  \var \Bigl[
  n_{A\uparrow}(r,\vartheta) -n_{A\downarrow}(r,\vartheta)\Bigr] \nonumber \\
  & = \tilde n^2 (\Re Q)^2 \frac{n_A}{\pi w^2}.
\end{align}

This motivates the definitions
\begin{align}
  \label{eq:10}
  N_{\uparrow,\downarrow} & \equiv \intlimits_0^{2 \pi}
  \intlimits_0^\infty \mathrm{d}\vartheta r \mathrm{d}r \,
  I_{P}(r) (\pi w^2 n_{A\uparrow}(r,\vartheta)), \nonumber \\
  N_A &\equiv\pi w^2 n_A, \quad n \equiv 2 (n_R+ t n_P), \quad k
  \equiv \frac{ \tilde n \Re Q} { n \pi w^2}
\end{align}
so that from Eq.~\eqref{eq:5} we can derive Eq.\eqref{eq:ja_2}: $
\phi = \frac{\Delta n}{n} = k \Delta N$. We find consistently
that for all atoms in the state~$\ket{\uparrow}$:
$\phi_\ket{\uparrow} = k N_\uparrow = k N_A$ and for the CSS:
$\langle \phi_\text{CSS}\rangle=0$, $\var(\phi_\text{CSS}) = k^2
\var(\Delta N) = k^2 N_A.$

\begin{acknowledgments}
  The work was funded by the Danish National Research Foundation, by
  EU grants COMPAS, EMALI, and QAP. N.K.  gratefully acknowledges
  support from the Danish Natural Science Research Council via a Steno
  Fellowship.  We thank R.~Le Targat and J.H.~M\"{u}ller for helpful
  discussions.
\end{acknowledgments}

{\footnotesize

\begin{thebibliography}{10}

\bibitem{WINELAND1992}
Wineland DJ, Bollinger JJ, Itano WM, Moore FL, Heinzen DJ
\newblock (1992) Spin squeezing and reduced quantum noise in spectroscopy.
\newblock {\em Phys. Rev. A} 46:R6797--R6800.

\bibitem{Kitagawa1993}
Kitagawa M, Ueda M
\newblock (1993) Squeezed spin states.
\newblock {\em Phys. Rev. A} 47:5138--5143.

\bibitem{Sorensen2001}
S{\o}rensen A, Duan LM, Cirac JI, Zoller P
\newblock (2001) Many-particle entanglement with {Bose}-{Einstein} condensates.
\newblock {\em Nature} 409:63--66.

\bibitem{Leibfried2004y}
Leibfried D, {\em et~al.}
\newblock (2004) Toward {H}eisenberg-limited spectroscopy with multiparticle
  entangled states.
\newblock {\em Science} 304:1476--1478.

\bibitem{Roos2006}
Roos CF, Chwalla M, Kim K, Riebe M, Blatt R
\newblock (2006) `{Designer} atoms' for quantum metrology.
\newblock {\em Nature} 443:316--319.

\bibitem{Hald1999}
Hald J, S{\o}rensen JL, Schori C, Polzik ES
\newblock (1999) Spin squeezed atoms: A macroscopic entangled ensemble created
  by light.
\newblock {\em Phys. Rev. Lett.} 83:1319--1322.

\bibitem{Kuzmich2000}
Kuzmich A, Mandel L, Bigelow NP
\newblock (2000) Generation of spin squeezing via continuous quantum
  nondemolition measurement.
\newblock {\em Phys. Rev. Lett.} 85:1594--1597.

\bibitem{Orzel2001}
Orzel C, Tuchman AK, Fenselau ML, Yasuda M, Kasevich MA
\newblock (2001) Squeezed states in a bose-einstein condensate.
\newblock {\em Science} 291:2386--2389.

\bibitem{Molmer2001}
S{\o}rensen A, M{\o}lmer K
\newblock (2001) Entanglement and extreme spin squeezing.
\newblock {\em Phys Rev. Lett.} 86:4431--4434.

\bibitem{Andre2004}
Andre A, S{\o}rensen A, Lukin M
\newblock (2004) Stability of atomic clocks based on entangled atoms.
\newblock {\em Phys Rev. Lett.} 92:230801.

\bibitem{Gerbier2006}
Gerbier F, F\"{o}lling S, Widera A, Mandel O, Bloch I
\newblock (2006) Probing number squeezing of ultracold atoms across the
  superfluid-{M}ott insulator transition.
\newblock {\em Phys. Rev. Lett.} 96:090401.

\bibitem{Fernholz2008}
Fernholz T, {\em et~al.}
\newblock (2008) Spin squeezing of atomic ensembles via nuclear-electronic spin
  entanglement.
\newblock {\em Phys. Rev. Lett.} 101:073601.

\bibitem{Takano}
Takano T, {\em et~al.}
\newblock (2009) Spin squeezing of  a cold atomic ensemble with the nuclear spin of one-half.
\newblock {\em Phys. Rev. Lett.} 102:033601.

\bibitem{Esteve2008}
Est\`eve J, Gross C, Weller A, Giovanazzi S, Oberthaler MK
\newblock (2008) Squeezing and entanglement in a {Bose}-{Einstein} condensate.
\newblock {\em Nature} 455:1216--1219.

\bibitem{Guerlin2007}
Guerlin C, {\em et~al.}
\newblock (2007) Progressive field-state collapse and quantum non-demolition
  photon counting.
\newblock {\em Nature} 448:889--893.

\bibitem{Caves1980}
Caves CM, Thorne KS, Drever RWP, Sandberg VD, Zimmermann M
\newblock (1980) On the measurement of a weak classical force coupled to a
  quantum-mechanical oscillator. {I}. issues of principle.
\newblock {\em Rev. Mod. Phys.} 52:341--392.

\bibitem{Atature2007}
Atature M, Dreiser J, Badolato A, Imamoglu A
\newblock (2007) Observation of {F}araday rotation from a single confined spin.
\newblock {\em Nature Physics} 3:101--105.

\bibitem{Santarelli1999}
Santarelli, G {\em et~al.}
\newblock (1999) Quantum projection noise in an atomic fountain: A high
  stability cesium frequency standard.
\newblock {\em Phys. Rev. Lett.} 82:4619--4622.


\bibitem{Kuzmich1998}
Kuzmich A, Bigelow NP, Mandel L
\newblock (1998) Atomic quantum non-demolition measurements and squeezing.
\newblock {\em Europhys. Lett.} 42:481--486.

\bibitem{Saffman2008}
Saffman M, Oblak D, Appel J, Polzik ES
\newblock (2008) Spin squeezing of atomic ensembles by multi-color quantum
  non-demolition measurements.
\newblock {\em arXiv:0808.0516}.

\bibitem{Oblak2008}
Oblak D {\em et~al.}
\newblock (2008) Echo spectroscopy of atomic dynamics in a gaussian trap via
  phase imprints.
\newblock {\em Eur. Phys. J. D} 50:67-73 .

\bibitem{Windpassinger2008b}
Windpassinger PJ, {\em et~al.}
\newblock (2008) Inhomogeneous light shift effects on atomic quantum state
  evolution in non-destructive measurements.
\newblock {\em New J. Phys.} 10:053032.

\bibitem{Oblak2005}
Oblak D, {et~al.}
\newblock (2005) Quantum-noise-limited interferometric measurement of atomic
  noise: Towards spin squeezing on the Cs clock transition.
\newblock {\em Phys. Rev. A} 71:043807.

\bibitem{Windpassinger2008a}
Windpassinger PJ, {\em et~al.}
\newblock (2008) Nondestructive probing of rabi oscillations on the cesium
  clock transition near the standard quantum limit.
\newblock {\em Phys. Rev. Lett.} 100:103601.

\bibitem{Ozeri}
Ozeri R, {\em et~al.}
\newblock (2005) Hyperfine coherence in the presence of spontaneous photon scattering.
\newblock {\em Phys. Rev. Lett.} 95:030403.

\bibitem{Hammerer2008}
Hammerer K, S{\o}rensen A, Polzik ES
\newblock (2008) Quantum interface between light and atomic ensembles.
\newblock {\em arXiv:0807.3358}.




\bibitem{Meiser2008}
Meiser D, Ye J, Holland M
\newblock (2008) Spin squeezing in optical lattice clocks via lattice-based
  {QND} measurements.
\newblock {\em New J. Phys.} 10:073014.

\bibitem{Wilpers:2002}
Wilpers G, {\em et~al.}
\newblock ({2002}) {Optical clock with ultracold neutral atoms}.
\newblock {\em Phys.Rev.Lett.} {89}:230801 .

\bibitem{Ludlow:2008y}
Ludlow AD {\em et~al.}
\newblock (2008) Sr lattice clock at $10^{-16}$ fractional uncertainty by
  remote optical evaluation with a {Ca} clock.
\newblock {\em Science} 319:1805--1808.

\bibitem{Lodewyck}
Lodewyck J, Westergaard PG, Lemonde P. Non-destructive measurement of the transition probability in a Sr optical lattice clock. quant-ph:0902.2905

\bibitem{SchleierSmith2008}
Schleier-Smith MH, Leroux ID, Vuleti{\'c} V
\newblock (2008) Reduced-quantum-uncertainty states for an atomic clock.
\newblock {\em arXiv:0810.2582v1}.

\bibitem{Hart:2007}
Hart RA, Xu X, Legere R, Gibble K
\newblock ({2007}) {A quantum scattering interferometer}.
\newblock {\em Nature} {446}:{892--895}.
\end{thebibliography}
}

\end{article}

\newpage 
\begin{centering}
  \section{Supplementary Information}
  \section{\Large Mesoscopic atomic entanglement for
precision measurements beyond the standard quantum limit}
\end{centering}

\setcounter{figure}{0} \setcounter{equation}{0}

\renewcommand{\thefigure}{S\arabic{figure}}
\renewcommand{\theequation}{S\arabic{equation}}

\subsection{Verification of the Projection Noise level}

We perform projection noise measurements on a CSS to obtain
fluctuating phase shifts $\phi_\text{CSS}$ with $\var(\phi_\text{CSS})
= k^2 \var(\Delta N) = k^2 N_A$. After each measurement we pump all
the atoms into the $\ket{\uparrow}$ state and determine the phase
shift $\phi_\ket{\uparrow}= k N_\uparrow = k N_A$.  These relations
allow us to determine the maximum effective number of atoms from the
maximum phase shift observed $\phi_{\ket{\uparrow},\text{max}}$:
\begin{equation}
  \label{eq:14}
  N_{A,\text{max}} = \phi_{\ket{\uparrow},\text{max}}\cdot  \left( \frac{d \var(\phi_\text{CSS})}{d \phi_\ket{\uparrow}} \right)^{-1} = 1.2 \cdot 10^5.
\end{equation}

An independent way to determine $N_A$ is described as follows: We
determine $Q$ from
\begin{eqnarray}
  \label{eq:8}
  Q_\uparrow & = & - \frac{3 \lambda^2}{4 \pi} \gamma \left[
    \frac{5/9}{\Delta_\uparrow +i\gamma/2}+
    \frac{1/9}{\Delta_\uparrow+\unit[452]{MHz} +i \gamma/2} \right]
\end{eqnarray} and by inserting the probe beam radius $w$ into definition given in Eq. [14] of the main text
together with Eq. [7]
the maximum effective atom number can be deduced
from the maximum phase shift
$\phi_{\ket{\uparrow},\text{max}}=\unit[0.18]{rad}$: \begin{equation}
  \label{eq:11}
  N_{A,\text{max}} = \phi_{\ket{\uparrow},\text{max}} \cdot \frac{2 \pi w^2}{\Re Q_\uparrow} =1.8 \cdot 10^5.
\end{equation}

The values obtained from eq.~\eqref{eq:14} and eq.~\eqref{eq:11} show
reasonable agreement, considering that the approximation that the
atomic cloud is wide compared to the probe beam but its axial
extension is small compared to the probe beam Rayleigh length are only
approximately fulfilled.  The projection noise slope $\frac{d
  \var(\phi_\text{CSS})}{d \phi_\ket{\uparrow}}$ obtained in different
experimental runs over the course of several weeks shows a
reproducibility to within $\pm 10\%$.

\subsection{Consistency check of the atom number estimation}
Let $\alpha_{\uparrow\downarrow,\text{CSS}}$ denote the absorption
coefficient of a cloud of atoms, prepared in the CSS for the two probe
wavelengths.  The overall absorption coefficient for a dichromatic
light pulse can then be expressed as
\begin{equation}
  \label{eq:13}
  \alpha_\text{CSS} = \frac{\alpha_{\uparrow,\text{CSS}} +
    \alpha_{\downarrow,\text{CSS}}}{2} = n_A \frac{\Im Q_\uparrow + \Im Q_\downarrow}{4}.
\end{equation}

A light pulse with $2 n_P$ photons interacting with the atoms causes
$\alpha_\text{CSS} \cdot 2 n_P $ scattering events. Each photon scattered off an atom that is still in the superposition
state reveals this atom's internal state and therefore ruins the superposition.
Thus for each scattered photon per unit area the density
$n_\text{CSS}(r,\vartheta)$ of atoms that still reside coherently in
the $(\ket{\uparrow}+\ket{\downarrow})/\sqrt{2}$ superposition
is reduced by $n_\text{CSS}/n_A$. Hence for a pulse with the photon column
density $2 n_P I_P(r)$ one can derive the differential equation
\begin{equation}
  \label{eq:17}
  \textrm{d} n_\text{CSS} = - \alpha_\text{CSS} \frac{n_\text{CSS}}{n_A} I_P 2 \, \textrm{d} n_P,
\end{equation}
which has the solution
\begin{equation}
  \label{eq:18}
  n_\text{CSS}(n_P) = n_A \exp\left(- \alpha_\text{CSS} I_P \frac{2 n_P}{n_A}\right).
\end{equation}

The last step of the spin-echo sequence outlined above is the
measurement of the decoherence parameter $\eta=1-N_\text{CSS}/N_A$ by
determining the fraction of atoms that still form a CSS, weighted with
the probe beam profile $I_P(r)$:

\begin{equation}
  \label{eq:21}
  \eta = 1- \frac{N_\text{CSS}}{N_A} = 1-  \intlimits_0^{2 \pi}
  \intlimits_0^\infty \mathrm{d}\vartheta r \mathrm{d}r \,
  I_P(r) \frac{n_\text{CSS}}{n_A}
  = 1-  \intlimits_0^{2 \pi}
  \intlimits_0^\infty \mathrm{d}\vartheta r \mathrm{d}r \,
  I_P(r) \exp \left( - 2 n_P I_P(r) \frac{\Im Q_\uparrow + \Im Q_\downarrow}{4}\right).
\end{equation}

Using $w=\unit[27]{\mu m}$, $2 n_P=7.4 \cdot 10^6$ for our probe
frequencies eq.~\eqref{eq:21} predicts $\eta = 17\%$. This estimate
becomes closer to the the observed value of $11\%$ when optical losses induced by the vacuum
cell windows and the spatial profile of the atomic density within the
probe volume are taken into account.

\newpage
\subsection{The clock operation and the effect of spontaneous emission
  on the dichromatic QND measurement}

\begin{figure*}[b!]{
  \begin{center}
    \includegraphics[width=\columnwidth]{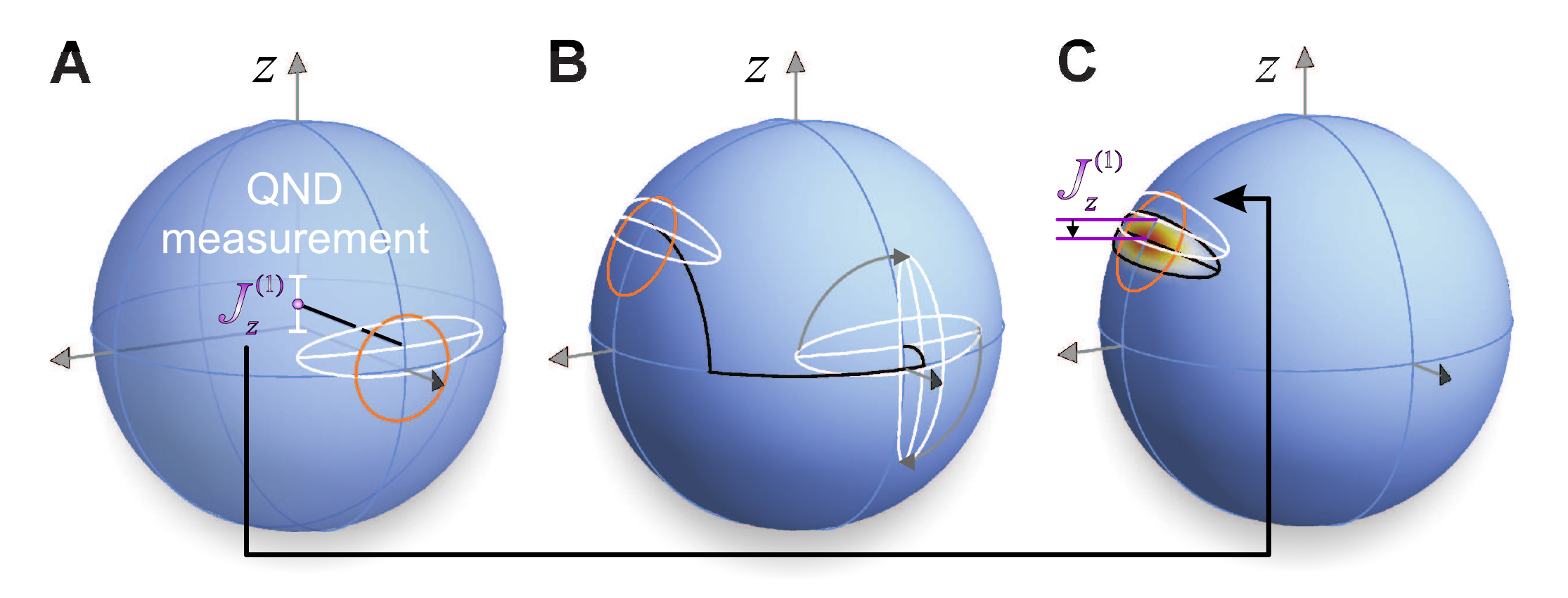}
  \end{center}
  \caption{(A) A QND measurement of $J_{z}$ with result $J_{z}^{(1)}$ creates an SSS displaced from the equator. (B) The
    SSS is rotated about the y axis and the Ramsey sequence proceeds
    as in Fig.~1A in the main paper. (C) The projection noise uncertainty of
    the clock signal can be reduced by using the knowledge of
    $J_{z}^{(1)}$.
    \label{fig:sub_one}}
  }
\end{figure*}

In this section we outline a Ramsey-sequence which would allow the
squeezed state created by the dichromatic QND measurement to improve
on the clock performance. This is essentially the sequence depicted
in Fig.~S1A-C. Consider a CSS pointing along the $y$-axis on the
Bloch sphere (Fig.~S1A). We perform a QND measurement of $J_z$ and
then apply the sequence S: $\pi /2$ -- rotation about $y$-axis,
$\varphi$ precession about $z$-axis, $\pi /2$ -- rotation about
$x$-axis.

This transformation corresponds to the unitary operator
$\hat{U}(\varphi )=\exp \left(-i\frac{\pi }{2} \hat{J}_{x} \right)\exp
\left(-i\varphi \hat{J}_{z} \right)\exp \left(i\frac{\pi }{2}
  \hat{J}_{y} \right)$. In the Heisenberg picture a final $J_z$
measurement performed after completion of the S sequence can
be related to the initial $J_z$ operator as follows:

\begin{equation}
  \label{eq:Jzinv}
  \hat{J}_{z}^\text{final} (\varphi)=\hat{U}^{\dag } (\varphi )\hat{J}_{z}^\text{initial}
  \hat{U}(\varphi )=\cos \varphi \hat{J}_{y}^\text{initial} -\sin
  \varphi \hat{J}_{z}^\text{initial}
\end{equation}
so that the expectation value $\left\langle \hat{J}_{z}^\text{final}
  (\varphi )\right\rangle =\cos \varphi \left\langle
  \hat{J}_{y}^\text{initial} \right\rangle $ can be used to determine
the precession angle $\varphi $. The highest sensitivity to changes of
$\varphi $ is achieved when $\varphi =-\pi/2$.  For this angle,
however, the projection noise $\var \, \left(\hat{J}_{z}^\text{final}
  (\varphi )\right)=\sin \varphi \,
\var\left(\hat{J}_{z}^\text{initial} \right)$ is also maximal. By
making use of the QND measurement of $J_{z}^\text{initial} $ prior to
the sequence S we can reduce these fluctuations conditionally. For
$\varphi =-\pi /2$ equation \eqref{eq:Jzinv} reduces to
$\hat{J}_{z}^\text{final} (-\pi /2)=\hat{J}_{z}^\text{initial} $ which
means that the spin projection along the $z$-axis is invariant under
this sequence and the outcome of a measurement of $J_{z}^\text{final}$
after the sequence S can be predicted up to the uncertainty following
the QND measurement of $J_{z}^\text{initial}$.

The invariance of $J_{z} $ in the proposed clock sequence has an
important implication for the influence of the atoms that have been
decohered by spontaneous scattering of photons from the QND probe
pulse. The key point is that in our scheme, each of the two probes
couples to one ground level population only: The selection rules
restrict spontaneously scattered atoms to decay back to the hyperfine
level they were initially in. Since the two probes measure the
populations in the whole hyperfine manifolds, i.e. one detects
$N_{\uparrow } $($F=4$, all sublevels) while the other detects
$N_{\downarrow } $($F=3$, all sublevels), the measured value of
$J_{z}={\tfrac{1}{2}} (N_{\uparrow } -N_{\downarrow } )$ is, to first
order, immune to spontaneous scattering.

Moreover, the clock sequence S only affects atoms in states $m_{F} =0$
because atoms in all other states are not coupled to the microwave
field and thus do not perform any of the rotations. This isolation is
due to their resonant microwave frequencies being shifted with respect
to the $F=4,m_{F} =0\leftrightarrow F=3,m_{F} =0$ transition frequency
by a bias magnetic field. Therefore the contribution of these atoms to
the measured $J_{z} $ is also invariant under the clock sequence.  The
conclusion is that the total $J_{z} $ measured by the dichromatic QND
method is invariant under the clock sequence S irrespectively of the
spontaneous emission, and hence its contribution to the noise of the
clock measurement can be effectively reduced by this approach.

A deviation from the above description comes from the fact that the
$\pi $-polarized probes couple slightly differently to $m_{F} =0$ and
$m_{F} =\pm 1$ levels. This adds some small extra noise to the spin
squeezed state, which in principle could even be avoided by repumping
all atoms with $m_F\neq 0$ into the according $m_F=0$ state.


\end{document}